\documentstyle[aps,epsfig,multicol]{revtex} 

\newcommand{\be}{\begin{equation}}
\newcommand{\ee}{\end{equation}}
\newcommand{\bea}{\begin{eqnarray}}
\newcommand{\eea}{\end{eqnarray}}

\newcommand{\lp}{\left(}
\newcommand{\rp}{\right)}

\begin{document}

\twocolumn[\hsize\textwidth\columnwidth\hsize\csname @twocolumnfalse\endcsname
\title{Quantum spin chains and Majorana states in arrays of coupled qubits}
\author{L. S. Levitov$^1$, T. P. Orlando$^2$, J. B. Majer$^3$, J. E. Mooij$^3$}

\address{$(1)$ Physics Department and Center for Materials Science \& Engineering, 
Massachusetts Institute of Technology, 77 Massachusetts Ave, Cambridge, MA 02139\\
$(2)$ Electrical Engineering Department, Massachusetts Institute of Technology, 77 Massachusetts Ave, Cambridge, MA 02139\\
$(3)$ Applied Physics and DIMES, Delft Technical University, 
P.O.Box 5046, 2600 GA Delft, the Netherlands}
%\date{\today}

\maketitle

\begin{abstract}
Several designs of inter-qubit coupling are considered. It is shown that by 
a combination of Josephson and capacitive coupling one can realize 
qubit interactions of variable spin content. Qubit arrays 
are discussed as models of quantum spin chains. In particular,
a qubit model of the 1D quantum Ising spin chain is proposed. 
A realization of unpaired Majorana fermion states in this system 
is considered. It is shown that Majorana states are represented by spin flip 
excitations localized on the chain ends. Using unpaired Majorana states 
in qubit chains for decoherence protected quantum computing is discussed.
\vskip2mm
\end{abstract}
]
\bigskip 

%\begin{multicols}{2}
\narrowtext

%%%%%%%%%%%%%%%%%%%%%%%%%%%%%%%%%%%%%%%%%%%%%%%%%%%%%%%%%%%%%%%%%%%%

Quantum computers are machines  that store information on quantum variables 
and that process that information by making those variables interact in a way
that preserves quantum coherence\cite{Divincenzo,Lloyd93}. 
Quantum superconducting circuits of nanoscale size became 
available recently for experiments which investigate 
novel quantum phenomena and their application to quantum 
computing\cite{Mooij99,Orlando99,Bocko97,Shnirman,Maklin99,Averin98,Nakamura99}.
This research has led to the observation of Rabi oscillations\cite{Nakamura99},
and to a demonstration of the superposition of macroscopic quantum states, the
so-called ``Schr\"{o}dinger cat'' states\cite{Caspar00,Friedman00}. 
The superposition character of these states was revealed\cite{Caspar00} 
by doing microwave spectroscopy experiments on the 
two quantum levels of superconducting circuits consisting of  three Josephson
junctions comprising the so so-called persistent current qubit
\cite{Mooij99,Orlando99}.
The two states of this qubit
have persistent currents of about a  microamp and correspond
to the  motion of millions of electrons. 

In this article we discuss circuits consisting of many coupled qubits. 
As a simplest example of such a circuit we consider an array of identical 
qubits with nearest neighbor couplings. 
Our interest in such circuits is two-fold. Firstly, qubit arrays represent a system
in which novel types of qubit couplings with different `spin operator 
content' and tunable control over the coupling strength can be investigated. 
Secondly, such arrays can be used to explore novel quantum phenomena with 
possible applications in quantum computing. 

Even the simplest 
of these systems discussed below, the so-called `quantum Ising spin chain,' 
is rather rich. Several interesting many-body states can be realized 
by varying qubit parameters and their coupling strength. Elementary excitations
in some of these states are quantum solitons
%% \ftn{discuss application in quantum information?} 
theoretically described as 
pseudofermions. We consider the phase diagram and point out several 
interesting regions in which the problem is exactly solvable. 
Also, as discussed below, the so-called Majorana fermion states can be 
realized as midgap states localized at the ends of the quantum Ising chain. 
In a recent article \cite{Kitaev} Kitaev outlined the possibility 
of using such states for reducing environmental 
sensitivity of quantum computers. We show that the coupling of 
Majorana states to external time varying field can be weaker than that
of a single qubit.

%% \ftn{the novel quantum phenomena with
%% unconventional types of quantum `spin' chains which are difficult to
%% realize with one dimensional magnetic materials but which can be
%% designed with tunable parameters by using superconducting quantum
%% circuits} 

\begin{figure}[h]
%\vspace{5mm}
\centerline{\psfig{file=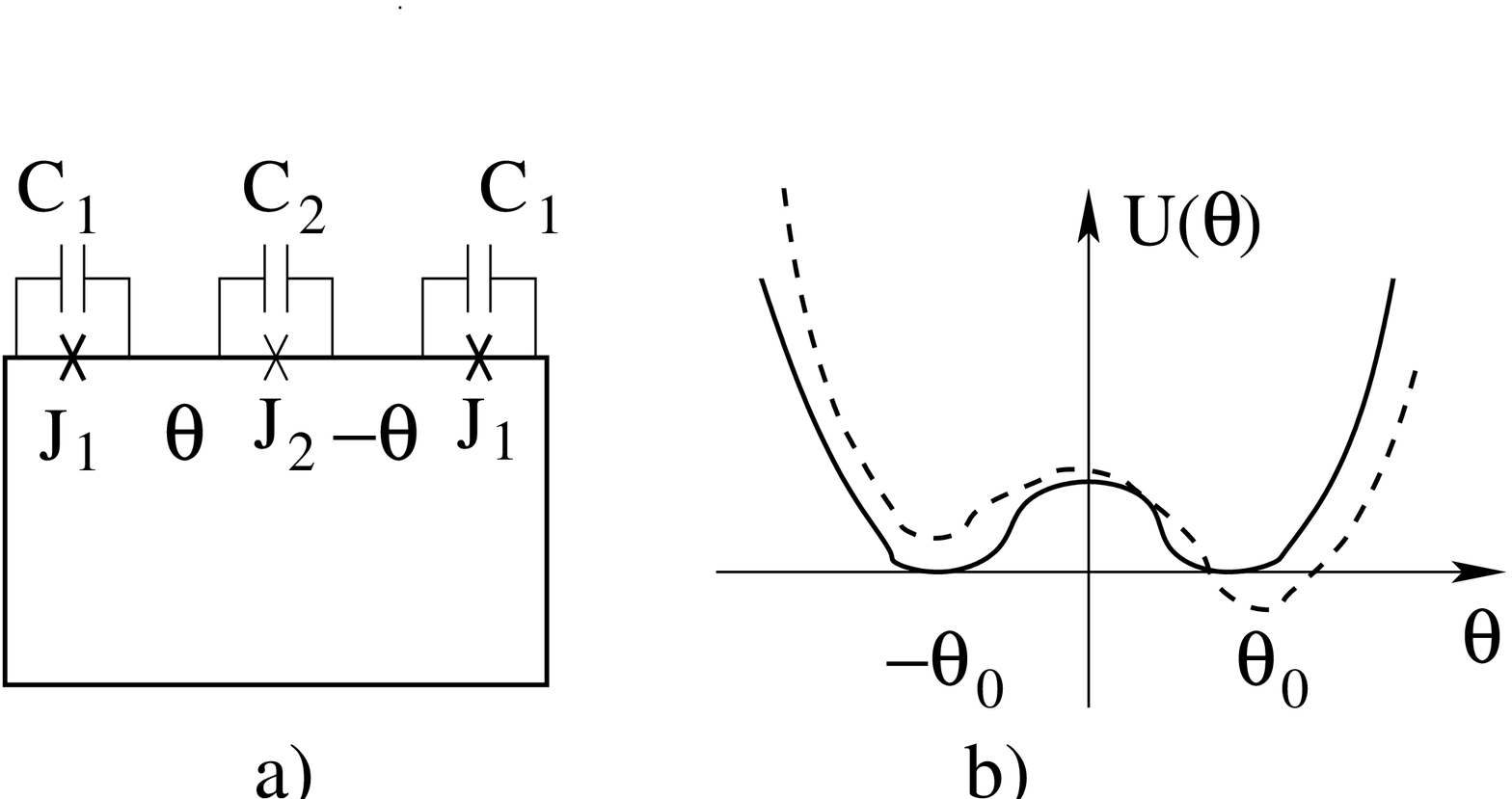,width=3.5in}}
\vspace{5mm}
%\epsfxsize=2.75in
%\epsfbox{fig1.eps}
 \caption[]{a) Qubit schematic shown with phases $\pm\theta$ 
on superconducting islands.
b) Josephson energy of the system as a function of the phase $\theta$. Solid
line: unbiased system with $\Phi=0.5\Phi_0$; dashed line: system with 
finite detuning $\delta\phi=\Phi- 0.5\Phi_0$.
     }
\label{LL1_fig}
  \end{figure}

The persistent current qubit \cite{Mooij99,Orlando99}
comprises a superconducting loop interrupted by
three Josephson junctions (marked in Fig.\ref{LL1_fig}(a) by an `$\times$') 
with the junction capacitance $C_{1,2}$.
The values of the three Josephson junctions
coupling constants $J_1$ and $J_2$ are chosen so that
the Josephson part of the Hamiltonian alone defines a bistable system
which at the value of external magnetic flux $\Phi=0.5\Phi_0$
can be either in the right-hand, or in the
left-hand current state.
In this qubit, by choosing parameters
$J_{1,2}$ and $C_{1,2}$ appropriately,
the barrier in the phase space separating the
right and left current states can be made low enough, 
so that tunneling between two classical states will take place. 
Also, detuning of external field from the value $0.5\Phi_0$
produces a bias on the right and left states, making one of them lower in 
energy than the other. 

%%%%%%%%%%%%%%%%%%%%%%%%%%%%%%%%%%%%%%%%%%%%%%%%%%%%%%%%%%%%%%%%%%%
%%%%%%%%%%%%%%%%%%%%%%%%%%%%%%%%%%%%%%%%%%%%%%%%%%%%%%%%%%%%%%%%%%%

It was shown in Ref. \cite{Mooij99} that
the phase space coordinate which describes transitions between the
right and left states is the relative phase $\theta$ of the 
two superconducting islands [Fig.\ref{LL1_fig}(a)]. 
The potential energy of the system as a function of $\theta$
has two minima, $\theta=\pm\theta_0$, as shown in Fig.~\ref{LL1_fig}(b).
These minima are symmetric at $\Phi=0.5\Phi_0$ and asymmetric 
at $\Phi\ne 0.5\Phi_0$. Tunneling takes place through the barrier
separating the two minima. The resulting qubit Hamiltonian is
  \begin{equation}\label{H1}
{\cal H}= -\Delta \sigma^x - h \sigma^z
\ ,
  \end{equation}
where $\Delta$ is the tunneling amplitude for the barrier $U(\theta)$,
and $h=2 I_p(\Phi-0.5\Phi_0)$  is proportional to the 
detuning $\Phi-0.5\Phi_0$, and where $I_p$ is the circulating current.
The  energy of the ground state and the first excited states
are given by 
  \begin{equation}\label{ERG1}
   E_{\mp} = \mp \sqrt{ h^2 + \Delta^2
   }
   \end{equation}
In the notation of Eq.~(\ref{H1}) the right and left current states
correspond to qubit `spin' up and down.
The energy levels repulsion manifest in Eq.~(\ref{ERG1}) was directly observed
recently by a spectroscopic technique\cite{Caspar00,Friedman00}.

The goal of 
this paper is to discuss a possibility to 
use superconducting quantum qubits to 
study novel quantum phenomena, control of qubit systems,
and novel quantum devices.
The question of how to make inter--qubit
couplings of required strength, sign, and spin content
is central for constructing logical gates out of 
qubits. We consider several basic couplings,
such as $zz$, $xx$, etc., in one-dimensional qubit arrays coupled
inductively, capacitively, or by additional Josephson junctions. 
One-dimensional geometry is relatively 
simple to manufacture, as well as it is simplest to work 
on theoretically in order to have theory
predictions compared with experimental results.
It is of interest to use arrays of coupled qubits to study
novel quantum phenomena which can not be simulated either with
present materials or present classical computers. 
In particular, qubit arrays can be used as models of 
quantum spin chains, one of the most basic systems 
in many-body physics, in parameter regimes that are not 
accessible with magnetic materials. 

\section{Inter--Qubit Couplings; Quantum Ising Problem}

Designing couplings is essential for being able to assemble qubits
into circuits capable of performing computation. For that, inter-qubit
couplings must satisfy several requirements. First, the coupling
strength should not be much less than the Rabi frequency of an
individual qubit, so that the states of different qubits entangle with
each other sufficiently quickly. Second, the coupling should be much
weaker than the Josephson plasma frequency which sets the scale for the
energy gap between the  two lowest energy qubit states and states with
higher energy. Third, the design should allow for different kinds of
coupling, such as purely transverse, purely longitudinal, or more
general couplings, as required by the particular scheme of processing
quantum bits. Fourth, it should be possible to turn the couplings on
and off on sufficiently short time scale. Below we outline our
approach to handling the problem of couplings.

Conceptually, the simplest kind of inter-qubit coupling is of the 
longitudinal $\sigma_1^z\sigma_2^z$ form, which can be easily
realized by magnetic interaction of qubits.  This coupling results
from magnetic field flux that the current in one qubit induces in the
other nearby qubit. The form of this coupling in terms of qubit
`spin' operators, since the right and left current states have
a definite $z$-component of qubit `spin,' is indeed
$\sigma_1^z\sigma_2^z$.  The sign of this coupling is such that it
favors antiparallel spin state of two coupled qubits.  However, the
numerical value of the inductive coupling estimated for conducting
loops of a few microns in size turns out to be at least an order of
magnitude smaller that typical values of the tunneling amplitude
$\Delta$ and detuning parameter $h$. Also, the 
inductive
coupling value is preset by the system layout. 
often the lack of tunability, as the coupling strength is
usually preset by the system layout.

\begin{figure}[h]
%\vspace{5mm}
\centerline{\psfig{file=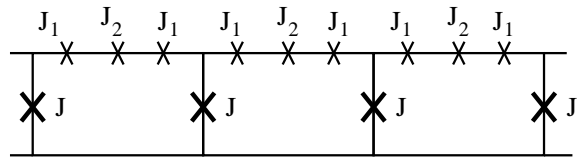,width=3in}}
\vspace{5mm}
%\epsfxsize=2.75in
%\epsfbox{fig1.eps}
 \caption[]{1D array of qubits coupled by shared Josephson
junctions: a realization of the $\sigma_1^z\sigma_2^z$ interaction. 
     }
\label{LL2_fig}
  \end{figure}

Because of that we propose another realization of the 
$\sigma_1^z\sigma_2^z$ coupling which has values tunable in a wider 
range than that of inductive coupling. 
The coupling is achieved by a Josephson junction
with a sufficiently large $J$ shared by adjacent qubits as shown in 
Fig.~\ref{LL2_fig}. In the regime when $J\gg J_{1,2}$ 
the phase drop across the large $J$ junction is much smaller than 
the overall phase change. As a result, this coupling is gentle enough
not to perturb the individual qubit dynamics.
However, the coupling strength $t$ is of the order 
of $J_{1,2}^2/J$, and thus it can be easily made of the same order as
the single qubit parameters $\Delta$ and $h$. The sign of this coupling 
is positive. 

%%%%%%%%%%%%%%%%%%%%%%%%%%%%%%%%%%%%%%%%%%%%%%%%%%%%%%%%%%%%%%%%%%%%
%%%%%%%%%%%%%%%%%%%%%%%%%%%%%%%%%%%%%%%%%%%%%%%%%%%%%%%%%%%%%%%%%%%%

As an example, we consider an array of qubits coupled 
by shared Josephson junctions, as shown in Fig.~\ref{LL2_fig}. The
Hamiltonian for this system is
  \begin{equation}\label{H2}
{\cal H}= \sum\limits_{i=-\infty}^{\infty} 
t\,\sigma_i^z\sigma_{i+1}^z
-(\Delta \sigma^x_i + h \sigma^z_i)
  \end{equation}
with positive $t$. Eq.~(\ref{H2}) describes one of the basic 
many-body systems: an antiferromagnetic spin $1/2$ chain 
with exchange constant $t$ 
in external magnetic field with components
$h$ and $\Delta$. We note that, since the inter-qubit coupling
is of the  $\sigma_1^z\sigma_2^z$ kind, the `exchange' interaction 
in the spin chain (\ref{H2}) is highly anisotropic. This should be contrasted
with nearly isotropic spin exchange coupling encountered in magnetic systems, 
with usually only weak anisotropy 
possible due to spin-orbital coupling. 

There are several reasons for considering the one-dimensional 
qubit array shown in Fig.~\ref{LL2_fig}.
One reason is that magnetic 
measurements on this system can provide a very direct
and simple test of inter-qubit couplings. To illustrate this point, 
we consider the system (\ref{H2}) in the 
limit $\Delta \ll h,\,t$. In this case the problem is reduced to
the classical 1D Ising model. Since the exchange coupling $t$ is positive, 
the ground state in the absence of external field is antiferromagnetic. 
Weak external field $|h|\ll t$ does not affect the ground state, while at high field all
spins align. 
Magnetization as a function of the field $h$ in this case is 
  \begin{equation}\label{M1}
\langle \sigma^z_i\rangle=\cases{1,\quad & for\ $h>2t$\cr
                                 0,\quad & for\ $|h|<2t$\cr
                                 -1,\quad & for\ $h<-2t$}
  \end{equation}
This should be compared with the magnetization step
reported recently for an array of isolated qubits \cite{Caspar00}. The result
(\ref{M1}) means that due to the qubit interaction (\ref{H2})
the magnetization step splits into two distinct steps with relative 
separation equal to $4t$. 
In a  real system the magnetization steps (\ref{M1})
can be smeared due to randomness in qubit parameters, finite 
value of $\Delta$, temperature, etc. Therefore, a demonstration of 
a magnetization curve similar to (\ref{M1}) would indicate high level 
of qubit reproducibility and control over qubit couplings. 

\begin{figure}[h]
%\vspace{5mm}
\centerline{\psfig{file=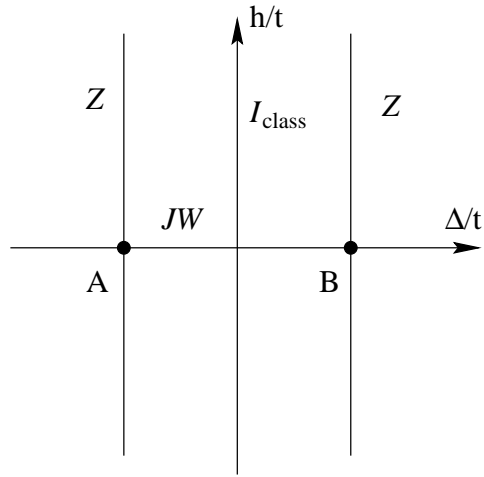,width=2.5in}}
\vspace{5mm}
%\epsfxsize=2.75in
%\epsfbox{fig1.eps}
 \caption[]{ The parameter space of the quantum Ising problem (\ref{H2}). 
On the line $I_{\rm class}$ the classical 1D Ising model is realized. 
On the line {\it JW} the problem is solved by 
the Jordan-Wigner transformation (\ref{JW-transform-Ising}),
and has fermion excitations. On this line the problem is equivalent to
the classical 2D Ising problem in the absence of external magnetic field, 
with the points $\Delta=\pm t$
corresponding to the 2D Ising critical point. On the lines {\it Z}
the problem is described by Zamolodchikov theory\cite{ZamolodchikovIsing} 
of critical Ising problem in magnetic field.
     }
\label{phase-diagram}
  \end{figure}

Another interesting fact is that the Hamiltonian (\ref{H2}) is nothing but
the quantum 1D Ising problem. Thus the qubit array shown in 
Fig.~\ref{LL2_fig} provides a physical model of this system.
The quantum 1D Ising problem is known to have nontrivial ordered
ground states and elementary excitations. The phase diagram 
of this system, sketched in Fig.\ref{phase-diagram}, can be easily obtained
by mapping the problem (\ref{H2})
onto the classical 2D Ising model with anisotropic couplings and 
associating the strong coupling axis with the time 
direction\cite{QuantumIsing}. 
For generic parameter values $h/t$ and $\Delta/t$ the problem 
is not integrable. 
However, on several lines indicated on the phase diagram there exist exact 
solutions. Different kinds of 
spin ordering in the 
ground state can be realized in this system. At zero `external field' $h$ 
there is (i) the ordered state with finite 
$\langle \sigma_i^z\rangle$ at $|\Delta|<t$, and (ii) the `disordered' 
state with finite $\langle \sigma_i^x\rangle$ at $|\Delta|>t$. 
The excitation spectrum 
has a gap at low energies which vanishes at the critical points 
$\Delta=\pm t$ (marked A and B in Fig.~\ref{phase-diagram}). 

It is instructive to consider the spin Hamiltonian (\ref{H2}) 
at the {\it JW} line $h=0$. In this case the problem
can be solved by the Jordan-Wigner transformation\cite{Fradkin,Tsvelik} 
\be\label{JW-transform-Ising}
\sigma^x_i=2a^\dagger_ia_i-1,\quad
\sigma^z_i=\lp a_i+a_i^\dagger\rp \prod\limits_{j<i}\sigma^x_j
\ee
The operators $a_i$ defined by (\ref{JW-transform-Ising}) 
describe fermions obeying standard
anticommutation relations, $\left\{a_i,a_j\right\}=0$,
$\left\{a_i^\dagger,a_j\right\}=\delta_{ij}$.
In terms of the fermions $a_i$, the Hamiltonian (\ref{H2}) reads
\be\label{Ising-fermi}
{\cal H}=\sum\limits_i 
t \lp a_i-a_i^\dagger\rp \lp a_{i+1}+a_{i+1}^\dagger\rp
-2\Delta a_i^\dagger a_i
\ee
The transformation (\ref{JW-transform-Ising}) and the resulting 
fermion problem (\ref{Ising-fermi}) are applicable to both 
finite and infinite spin chains.

%%%%%%%%%%%%%%%%%%%%%%%%%%%%%%%%%%%%%%%%%%%%%%%%%%%%%%%%%%%%%%%%%%%%
%%%%%%%%%%%%%%%%%%%%%%%%%%%%%%%%%%%%%%%%%%%%%%%%%%%%%%%%%%%%%%%%%%%%

We shall consider here the case of an infinite chain, when the 
Hamiltonian can be conveniently rewritten in the plane wave basis:
\be\label{Ising-fermi-fourier}
{\cal H}=\sum\limits_{p} 
t \lp a_p a_{-p} e^{-ip} + {\rm h.c.}\rp
%%% -a^\dagger_p a^\dagger_{-p} e^{ip}\rp 
- 2\lp \Delta + t \cos p \rp a^\dagger_p a_p
\ee
This problem is readily diagonalized by Bogoliubov transformation
yielding the dispersion relation of the form
\be\label{quantum-Ising-spectrum}
\epsilon(p)=2 \Large| \Delta + t\,e^{ip}\Large|,\quad
{\cal H}=\sum\limits_{p} \epsilon(p)\, b^\dagger_p b_p
\ee
with the excitation wavenumber in the Brillouin zone $-\pi<p<\pi$. 
The excitation spectrum (\ref{quantum-Ising-spectrum}) has a gap at low 
energies of the width 
$E_{\rm gap}=2\,{\rm min}\,\Large| \Delta\pm t\Large|$. 
The spectrum becomes gapless at $\Delta=\pm t$ (the critical points A and B
in the phase digram). 

Elementary excitations in this case are noninteracting Bogoliubov fermions. 
Let us remark that in terms of the original spin $1/2$ variables (\ref{H2})
an energy excitation can be viewed as consisting of 
independent solitons which propagate through the chain preserving 
their identity even when other excitations are present. 
This situation is familiar in integrable systems, classical and quantum. 
It may be of interest to consider using quantum solitons in
qubit arrays as means of quantum information transmission between 
distant parts of the qubit network. These solitons can in principle 
be generated and detected locally on individual qubits located at 
the opposite ends of the system. Such solitons can be used as 
information carrier in essentially the same way as solitons 
in nonlinear optical channels. 

\section{Majorana fermions in a qubit array}

In the theory of the quantum Ising model it is known\cite{GogolinTsvelik} 
that a rather
natural representation of the spin problem (\ref{H2}) can be obtained using
Majorana fermions defined in terms of the Jordan-Wigner fermions as
\be
c^1_j=\frac1{\sqrt{2}}\lp a_j+a_j^\dagger\rp,\quad
c^2_j=\frac1{\sqrt{2}i}\lp a_j^\dagger-a_j\rp
\ee
The operators $c^{1,2}_j$ are hermitian, $c^{a\dagger}_j=c^{a}_j$, 
and satisfy the anticommutation relations
\be
\left\{c^{a}_j c^{a'}_{j'} + c^{a'}_{j'} c^{a}_j\right\}
=\frac12 \delta_{aa'}\delta_{jj'},\quad
(c^{a}_j)^2=\frac12
\ee
The commutation relations indicate that each Majorana fermion represents
a `one-half' of a Jordan-Wigner fermion, so that the latter can be viewed 
as a pair of two Majorana fermions. 
It was proposed by Kitaev\cite{Kitaev} that
unpaired Majorana fermion states are
better protected from environmental noise than conventional fermions. 
Thus it is of interest to look for realizations of these states. 

Here we demonstrate that the qubit array shown in Fig.~\ref{LL2_fig} 
provides
under certain conditions a realization of unpaired 
Majorana fermions. 
The Hamiltonian (\ref{Ising-fermi}) for a chain of length $n$ 
can be rewritten in terms 
of Majorana fermions as
\be\label{Ising-majorana}
{\cal H}=2i\sum\limits_{j=1}^n 
t\, c^2_j c^1_{j+1}
-\Delta c^2_j c^1_j
\ee
The problem (\ref{Ising-majorana}) can be diagonalized by an orthogonal
$2n\times 2n$ Bogoliubov transformation of the operators $c^{1,2}_j$,
such that it brings the matrix
\be\label{matrix-majorana}
\left(
\matrix{0 & -\Delta & 0 & 0 & 0 &...\cr 
-\Delta & 0 & t & 0 & 0 & ... \cr
0 & t & 0 & -\Delta & 0 & ... \cr
0 & 0 & -\Delta & 0 & t & ... \cr
0 & 0 & 0 & t & 0 & ... \cr
... & ... & ... & ... & ... & ...}
\right)
\ee
to a diagonal form. The spectrum of this problem consists of two parts. 
Most of the states have energies above the gap $E_{\rm gap}$ found for 
the infinite qubit chain. In addition, there may exist two midgap states 
with energies within the gap near $\epsilon=0$. The midgap states are 
localized near the edges of the system. 

To illustrate this point, we consider a semi-infinite chain, and look for 
an eigenvector of the semi-infinite matrix (\ref{matrix-majorana})
with the eigenvalue $\epsilon=0$. This vector has the form
\be\label{midgap-vector}
A \left(
\matrix{1 \cr 0 \cr \lambda \cr 0 \cr \lambda^2 \cr ...}\right)
\ee
where $\lambda=\Delta/t$. For the state (\ref{midgap-vector}) to be 
normalizable, one must have $|\lambda|<1$, in which case 
the normalization factor is $A=(1-\lambda^2)^{-1/2}$. The condition
$|\lambda|<1$ indicates that the midgap states occur in the ordered 
Ising phase $t>|\Delta|$ (the AB interval in the phase diagram).

Note that only one Majorana species contribute to the midgap state, 
and so this is indeed an unpaired Majorana state.
For a finite but long chain
the midgap state on one end will be approximately given by 
(\ref{midgap-vector}), whereas the state on the other end of the 
chain will be defined in a similar way in terms of the other Majorana 
fermion species. For a finite chain the two midgap states will 
be coupled\cite{Kitaev}
by a matrix element of the order $\sim \lambda^n$. 

Let us discuss the meaning of these states in terms of the spin $1/2$ 
formulation (\ref{H2}). It is instructive to consider the limit 
$|\lambda|\ll1$, when the ground state of the spin chain is close 
to that of a classical antiferromagnet. In this case the width of the bulk 
gap is close to $2t$, which is consistent with the energy required to reverse
one spin in the bulk in the presence of coupling $t$ to the neighbors 
on both sides. 
An approximately twice smaller energy $t$ is required to reverse 
a spin at the end of the chain, which gives excitation energy 
in the middle of the bulk gap. 
Thus, each of the unpaired Majorana states at the ends of the chain 
is nothing but a single reversed spin, at finite $\lambda<1$ surrounded 
by slightly tilted spins on neighboring inner sites. 

%%%%%%%%%%%%%%%%%%%%%%%%%%%%%%%%%%%%%%%%%%%%%%%%%%%%%%%%%%%%%%%%%%%%
  \begin{figure}[h]
%\vspace{5mm}
\centerline{\psfig{file=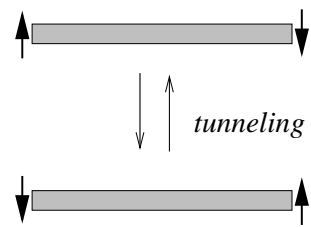,width=1.5in}}
\vspace{5mm}
%\epsfxsize=2.75in
%\epsfbox{fig1.eps}
 \caption[]{Tunneling between two unpaired Majorana excitations
realized as midgap states of the 1D quantum Ising problem.
     }
\label{fig-majorana}
  \end{figure}
%%%%%%%%%%%%%%%%%%%%%%%%%%%%%%%%%%%%%%%%%%%%%%%%%%%%%%%%%%%%%%%%%%%%

When the system is excited into one of the two midgap states, 
tunneling between two degenerate states will take place 
due to coupling of the midgap states on opposite ends across 
the chain. This tunneling process involves 
simultaneous coherent flipping of the spins at the chain
ends accompanied by rearrangement of the tilted spins near the 
ends.
%%% , as illustrated in Fig.~\ref{fig-majorana}.
It is noteworthy that this spin flipping does not change the net spin,
since the two spins at the ends flip coherently and in antiphase.
Hence the (spatially uniform) background magnetic field fluctuations 
on the two ends are cancelled out and do not affect the tunneling dynamics. 
This is consistent with the conclusion of Kitaev that unpaired 
Majorana states have weaker sensitivity to environmental time varying 
fields. 

\section{Transverse Inter--Qubit Couplings}

\begin{figure}[h]
%\vspace{5mm}
\centerline{\psfig{file=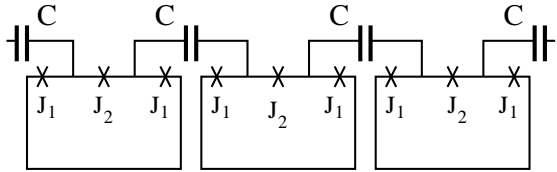,width=3in}}
\vspace{5mm}
%\epsfxsize=2.75in
%\epsfbox{fig1.eps}
 \caption[]{1D array of capacitively coupled
qubits: a realization of transverse inter-qubit coupling
described by the Hamiltonian (\ref{H3}).
     }
\label{LL3_fig}
  \end{figure}

Here we consider several interesting realizations of transverse couplings.
One possible kind of transverse
coupling is illustrated in Fig.~\ref{LL3_fig}. In this scheme, 
superconducting islands of adjacent qubits are capacitively
coupled in such a way that the coupling capacitance $C$ is of the order
of the Josephson junction capacitances $C_{1,2}$. The physical reason 
for the coupling shown in Fig.~\ref{LL3_fig} to be transverse is
the following. When qubits are in a superposition of the right and left 
current states, their quantum dynamics is described by tunneling 
between two minima of the potential $U(\theta)$ in Fig.~\ref{LL1_fig}. 
Each tunneling event produces a charge impulse on superconducting
islands. By making the qubits capacitively coupled, one enforces 
the correlation of time-dependent charge fluctuations on different 
islands, which is equivalent to having correlation of tunneling events on 
coupled qubits. The Hamiltonian for this coupling, as we shall see, 
is written in terms of the transverse spin operators 
$\sigma^{\pm}=\frac12(\sigma^x \pm i\sigma^y)$.

%%%%%%%%%%%%%%%%%%%%%%%%%%%%%%%%%%%%%%%%%%%%%%%%%%%%%%%%%%%%%%%%%%%%
%%%%%%%%%%%%%%%%%%%%%%%%%%%%%%%%%%%%%%%%%%%%%%%%%%%%%%%%%%%%%%%%%%%%

\begin{figure}[h]
%\vspace{5mm}
\centerline{\psfig{file=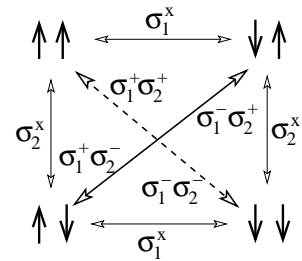,width=1.5in}}
\vspace{5mm}
%\epsfxsize=2.75in
%\epsfbox{fig1.eps}
 \caption[]{Correlated tunneling in two capacitively coupled qubits.
     }
\label{LL4_fig}
  \end{figure}

To derive the form of the  qubit coupling Hamiltonian, we consider two qubits
coupled capacitively according to the scheme
in Fig.~\ref{LL3_fig}. We represent the collective dynamics of the
two qubits by motion in the two-dimensional phase space parameterized 
by the qubits' phases $\theta_1$, $\theta_2$. Potential energy is given 
by the sum of Josephson contributions $U(\theta_1)+U(\theta_2)$. 
In the absence of qubit coupling, their
kinetic energy is $m(\dot\theta_1^2+\dot\theta_2^2)/2$ with 
effective mass $m$ being determined by the Josephson junction capacitances
$C_{1,2}$ --- see Ref.\cite{Mooij99}. Since the kinetic energy is isotropic,  
all tunneling amplitudes between four minima
$\theta_{1,2}=\pm\theta_0$ are equal to each other. In this case, 
as shown schematically in Fig.~\ref{LL4_fig}, tunneling of 
the two qubits is independent, and thus the tunneling Hamiltonian 
is written as a sum $\sigma^x_1+\sigma^x_2$. 
Now, the capacitive coupling shown in Fig.~\ref{LL3_fig} is described 
by adding to the kinetic energy a new term 
$m_\ast(\dot\theta_1+\dot\theta_2)^2/8$, where $m_\ast=(\hbar/2e)^2 C$. 
This term makes kinetic energy {\it anisotropic} by increasing effective mass
for the motion in the $(1,1)$ direction. As a result, when $m_\ast\ge m$, 
all tunneling amplitudes are suppressed with the exception of the
correlated transitions 
$|\!\uparrow\downarrow\rangle\leftrightarrow |\!\downarrow\uparrow\rangle$
illustrated in Fig.~\ref{LL4_fig}.
This process is described by the Hamiltonian of the form 
$\sigma^+_1\sigma^-_2+\sigma^-_1\sigma^+_2$. In the more general situation 
with $m_\ast\sim m$, both the correlated and independent tunneling processes 
take place, and the resulting Hamiltonian of the array
has the form
  \begin{equation}\label{H3}
{\cal H}= \sum\limits_{i=-\infty}^{\infty} 
t(\sigma_i^+\sigma_{i+1}^- +\sigma_i^-\sigma_{i+1}^+)
-(\Delta \sigma^x_i + h \sigma^z_i)
\ ,
  \end{equation}
where $\Delta$ is the individual qubit tunneling amplitude, and 
$t$ is the correlated tunneling amplitude. Note that $t\gg\Delta$ when 
$m_\ast\gg m$. 
The first term of Eq.~(\ref{H3}) can also be written as 
$(t/2)(\sigma_i^x\sigma_{i+1}^x +\sigma_i^y\sigma_{i+1}^y)$, i.e., 
the Hamiltonian (\ref{H3}) defines the so-called $XY$ spin model in 
an external field.

%%%%%%%%%%%%%%%%%%%%%%%%%%%%%%%%%%%%%%%%%%%%%%%%%%%%%%%%%%%%%%%%%%%%
%%%%%%%%%%%%%%%%%%%%%%%%%%%%%%%%%%%%%%%%%%%%%%%%%%%%%%%%%%%%%%%%%%%%

Magnetization curve can be computed at $\Delta\ll t$ by using
the Jordan--Wigner mapping of the spin $1/2$ problem to a free fermion 
problem. In this case, as we discuss below, the mapping
is constructed somewhat differently from the one used in the quantum 
Ising model. The result is:
  \begin{equation}\label{M2}
\langle \sigma^z_i\rangle=
\cases{1,\quad & for $h>t$\cr
       \frac2{\pi}\,{\rm arcsin}(h/t),\quad & for $|h|<t$\cr
       -1,\quad & for $h<-t$}
  \end{equation}
In the transverse coupling case, 
instead of two abrupt magnetization steps (\ref{M1}) 
found for the qubit array with the $zz$ coupling, we have a continuous 
magnetization curve with characteristic square root singularities
at $h=\pm t$. 

\begin{figure}[h]
%\vspace{5mm}
\centerline{\psfig{file=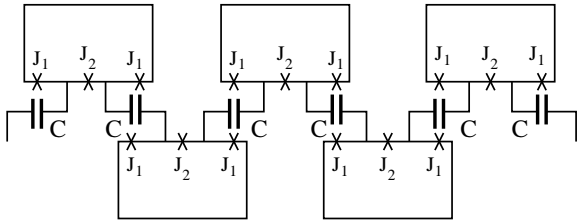,width=3in}}
\vspace{5mm}
%\epsfxsize=2.75in
%\epsfbox{fig1.eps}
 \caption[]{Another realization of an array with transverse couplings 
described by the Hamiltonian (\ref{H4}). 
     }
\label{LL5_fig}
  \end{figure}

To illustrate flexibility of the
systems with transverse coupling, we
consider an array with capacitive coupling built according to the 
scheme displayed in Fig.~\ref{LL5_fig}. Like in the array in 
Fig.~\ref{LL3_fig}, effective qubit `spin' couplings 
are purely transverse in this case. The analysis of couplings goes along the 
same lines as for the array in Fig.~\ref{LL3_fig}. Two decoupled qubits
can be described by separable dynamics in the two-dimensional phase space 
$(\theta_1,\theta_2)$. The coupling changes kinetic energy
by adding to it the term
$m_\ast(\dot\theta_1-\dot\theta_2)^2/2$, where $m_\ast=(\hbar/2e)^2C$.
Note that in this case, due to different orientation of coupled qubits,
their coupling depends on the difference of the phases $\theta_1-\theta_2$,
rather than on the sum as in the above example. As a result, effective mass
is higher for the motion in the $(1,-1)$ direction, and 
at $m_\ast\gg m$ all tunneling processes except 
$|\!\uparrow\uparrow\rangle\leftrightarrow |\!\downarrow\downarrow\rangle$
are suppressed. According to Fig.~\ref{LL4_fig}, this process is described 
by the Hamiltonian of the form
$\sigma^+_1\sigma^+_2+\sigma^-_1\sigma^-_2$. As before, for 
the more general situation 
of $m_\ast\sim m$ there are both the correlated and independent 
tunneling processes,
and thus the Hamiltonian of the array in Fig.~\ref{LL5_fig}
is 
  \begin{equation}\label{H4}
{\cal H}= \sum\limits_{i=-\infty}^{\infty} 
t(\sigma_i^+\sigma_{i+1}^+ +\sigma_i^-\sigma_{i+1}^-)
-(\Delta \sigma^x_i + h \sigma^z_i)
\ .
  \end{equation}
The first term of Eq.~(\ref{H4}) can also be written as 
$(t/2)(\sigma_i^x\sigma_{i+1}^x -\sigma_i^y\sigma_{i+1}^y)$.

%%%%%%%%%%%%%%%%%%%%%%%%%%%%%%%%%%%%%%%%%%%%%%%%%%%%%%%%%%%%%%%%%%%%
%%%%%%%%%%%%%%%%%%%%%%%%%%%%%%%%%%%%%%%%%%%%%%%%%%%%%%%%%%%%%%%%%%%%

The difference between the problems (\ref{H3}) and (\ref{H4})
is most clearly displayed in the Jordan--Wigner representation
%%%% \cite{Fradkin},
  \begin{equation}\label{JW}
\sigma_i^+=a_i^+\prod\limits_{j<i}\sigma_j^z
,\ \
\sigma_i^-=a_i\prod\limits_{j<i}\sigma_j^z
,\ \ 
\sigma_i^z=2a_i^+a_i-1
.
  \end{equation}
To simplify matters we shall limit ourselves to 
the strong coupling case, when $m_\ast\gg m$ and thus
$t\gg\Delta$ in both (\ref{H3}) and (\ref{H4}). Then the fermionic
representation of the Hamiltonians (\ref{H3}) and (\ref{H4})
is given by
  \bea\label{H5}
{\rm (i):}\ {\cal H}= \sum\limits_{i=-\infty}^{\infty} 
t(a_i^+a_{i+1} +a_{i+1}^+a_i)
- 2h a_i^+a_i
\ ,\quad \\
{\rm (ii):}\ {\cal H}'= \sum\limits_{i=-\infty}^{\infty} 
t(a_i^+a_{i+1}^+ +a_{i+1}a_i)
- 2h a_i^+a_i
\ .
  \eea
The spectrum of elementary excitations is obtained by
going to the Bloch plane waves representation (in the case (ii) 
accompanied by Bogoliubov transformation). The result reads
  \bea\label{E(k)}
{\rm (i):}\quad \epsilon(p)=2(t\cos(p)-h)
\ ,\qquad \\
{\rm (ii):}\quad
\epsilon(p)=\pm 2\sqrt{t^2\sin^2(p)+h^2}
\ .
  \eea
It follows from the result (\ref{E(k)}) that the ground states 
of the two systems
are described by collective entangled states of all qubits in the chain,
represented by ideal Fermi gas in the case (i)
and by Dirac vacuum in the case (ii). In the case (i) the excitation 
spectrum is gapless for not too large $|h|<t$, whereas in the case (ii)
there is a finite gap in the excitation spectrum equal to $4|h|$. 

\section{Novel Quantum Phenomena}

%%% Experimentally, 
%%% we plan to fabricate  such qubit arrays and measure
%%% magnetization curves. On the theoretical side, we plan to study the
%%% effects of the tunneling amplitude $\Delta$ as well as of 
%%% finite temperature and disorder.

As we discussed above, there are several attractive schemes
of designing qubit couplings. Besides allowing for control over coupling 
strength and its spin operator content, the proposed schemes 
are compatible with the plans to achieve tunability of couplings in time.
This can be attempted, for example, by employing switchable Josephson 
junctions similar to the gated $InGaAs$ superconducting 
junctions \cite{Akazaki96-switch} which have very short switching times 
of the order of a few microseconds. 
%% We plan to test various couplings using one-dimensional qubit arrays. 
%% One-dimensional geometry has the advantage of being 
%% relatively simple to manufacture and also simple enough to
%% work on theoretically. 
Magnetization measurements on these arrays
compared with theoretical results can provide useful 
information on qubit couplings. 

%%% \section{Novel Quantum Phenomena}

%%% We plan to use arrays and networks of coupled qubits to
%%% explore a number of many--body phenomena of current interest.
%%% Besides contributing to fundamental physics, this research direction
%%% should enhance our ability to manipulate, measure and control 
%%% qubit states, individually and collectively. Below we focus on the
The one-dimensional arrays discussed above 
%%% (see Figs \ref{LL2_fig}, \ref{LL3_fig}) 
provide novel 
realization of spin $1/2$ chains, one of the 
basic many-body physics models. 
Quantum spin chains are well studied theoretically and 
are known to be quite rich in properties \cite{Fradkin,Tsvelik}.
Depending on the kind of coupling they can display different kinds of
ordering:  ferromagnetic or antiferromagnetic (both with algebraic
correlations), spin liquid, Luttinger liquid, etc. 
Elementary excitations (quasiparticles)
in these systems can have fractional quantum numbers and fractional (anyon)
statistics. Depending on couplings, the excitation
spectrum can be gapless or gapped, which should be manifest
in dynamical properties, i.e. in the character of excitation
transport along the chain.

Experimentally, the only realization of spin chains that has been available 
so far is in quasi one-dimensional magnetic materials. That, however, 
corresponds to a relatively small domain in the coupling parameter 
space, because spin exchange is usually isotropic or nearly isotropic, 
as a result of the  weakness of spin-orbital interaction in solids. 
Also, dynamical phenomena
such as excitation transport are quite difficult to access in magnetic
systems because available experimental techniques are limited to 
low frequency magnetization measurement and polarized neutron scattering.  
Thus realizing spin chains in the qubit arrays looks attractive 
from both points of view. First, in the qubit array couplings can 
be tuned to the desired form by the methods outlined in the previous section.
Second, the dynamics can be directly probed by standard electric
measurements. 

One can study excitation transport in the qubit arrays. 
Experimentally, this can be achieved by exciting the chain at one end, 
and measuring response at the opposite end. It will be interesting 
to manufacture and compare properties of gapless and gapped
systems, as the character of excitation transport has to be very 
different in the two cases. Moreover, in the case of the 1D quantum Ising 
problem (\ref{H2}), the dependence of the excitation
energy (\ref{quantum-Ising-spectrum}) on the qubit coupling $t$ 
indicates that for the array shown in Fig.~\ref{LL2_fig} both regimes 
of excitation transport are available simultaneously. 
By changing the qubit coupling, one can sweep 
through the range of parameters in 
which the gap in the excitation spectrum (\ref{quantum-Ising-spectrum}) 
will gradually close and than reopen. 
The problem of excitation transport is also quite interesting 
theoretically. The challenge here is that, in contrast with 
other one-dimensional Fermi systems such electrons in quantum 
wires, generic Hamiltonians of  spin chains do not 
conserve the number of Jordan-Wigner fermions. 
For example, the external field term $-h\sigma^z_i$ in the 1D quantum Ising 
Hamiltonian (\ref{H2}) has a nonlocal form in the Jordan-Wigner 
representation. This may in principle change the properties 
of elementary excitations. 

%%% We plan to work on the theory of the {\bf microscopic mechanism} 
%%% of excitation transport at different temperatures
%%% and levels of disorder. A priori it is not clear 
%%% whether excitations propagate via coherent inter-qubit  
%%% tunneling or incoherent hopping.
%%% Understanding the character of excitation transport in qubit arrays
%%% will also shed light on the long-standing problem of {\it spin-diffusion}.
%%% The question of whether spin waves at finite temperature propagate 
%%% diffusively or ballistically has been a subject of controversy in 
%%% the literature and essentially remains open \cite{Narozhny98}. 

Looking somewhat ahead of experimental developments, it is quite 
interesting to imagine and explore the possibilities that will 
become available if excitations in qubit arrays can indeed propagate 
ballistically, as suggested by the dispersion relation 
(\ref{quantum-Ising-spectrum}). One can ask whether it is
possible to design qubit analogs of electron-based quantum wires 
and use them to manipulate excitations in the same way as it is done in 
conventional nanoscaled semiconductor structures. Doing this should include 
developing a `battery' (i.e., a DC spin 
current source), as well as techniques of measuring the 
current flux of spin excitations. This is of interest 
because spin excitations, being neutral, are practically decoupled from
external electric and (to a lesser extent) magnetic fields. Thus it may be 
possible to achieve levels of coherence unavailable in electron 
systems (but perhaps comparable to what is possible for 
photons in quantum optics devices). 
%%% If true, this will have a number 
%%% of important consequences for nanostructure physics. We plan to work 
%%% theoretically on the realization of these ideas. 

Another interesting direction is to use qubit arrays to design unpaired 
Majorana fermion states. As the above discussion reveals, these states 
can be realized in the quantum Ising chain, i.e., in a qubit array with 
$zz$ couplings. On the grounds of what has been discussed 
above, one can expect that these states can be probed spectroscopically 
in a way similar 
to individual qubit states\cite{Caspar00}. Using this technique one can, 
in principle, determine decoherence time of Rabi oscillations realized with 
two Majorana states and verify theoretical expectation that unpaired 
Majorana states are protected from decoherence.

%%% Another interesting direction is to realize {\bf spin higher than $1/2$} 
%%% using several coupled qubits.
%%% This is of interest because excitations in integer spin 
%%% systems are expected to obey Bose statistics, whereas in the 
%%% half-integer spin systems excitations are fermions, as discussed above.
%%% Realization of higher spin may also have application in quantum computing
%%% because it can be used to realized the so-called Haldane gap 
%%% state \cite{Fradkin}
%%% which has a topologically nontrivial ordering resulting in certain classes
%%% of excitations being topologically protected and therefore robust.
%%% To conclude this section, one can say that once quantum properties of qubit 
%%% arrays are demonstrated, it will be possible just to take a text on 
%%% one-dimensional quantum problems, like Refs. \cite{Fradkin,Tsvelik},
%%% and simply read it. Rich properties of one-dimensional field theories
%%% realized in nanostructures should inevitably bring completely new insights.

We are grateful to Alexei Tsvelik and Seth Lloyd for useful discussions.
This work was partially supported by the MRSEC Program 
of the  
National Science Foundation 
under Grant No. DMR 98-08941. 

%%%%%%%%%%%%%%%%%%%%%%%%%%%%%%%%%%%%%%%%%%%%%%%%%%%%%%%%%%%%%%%%%%%%

\end{document}